# Moon Search Algorithms for NASA's Dawn Mission to Asteroid Vesta

Nargess Memarsadeghi[*a], Lucy A. McFadden[a], David R. Skillman[a], Brian McLean[b], Max Mutchler[b], Uri Carsenty[c], Eric E. Palmer[d], and the Dawn Mission's Satellite Working Group

[a]NASA Goddard Space Flight Center, Greenbelt, Maryland, USA
[b]Space Telescope Science Institute, Baltimore, Maryland, USA
[c]DLR, Institute of the Planetary Research, Berlin, Germany
[d]Planetary Science Institute, Tucson, Arizona, USA

## ABSTRACT

A *moon* or *natural satellite* is a celestial body that orbits a planetary body such as a planet, dwarf planet, or an asteroid. Scientists seek understanding the origin and evolution of our solar system by studying moons of these bodies. Additionally, searches for satellites of planetary bodies can be important to protect the safety of a spacecraft as it approaches or orbits a planetary body. If a satellite of a celestial body is found, the mass of that body can also be calculated once its orbit is determined. Ensuring the Dawn spacecraft's safety on its mission to the asteroid (4) Vesta primarily motivated the work of Dawn's Satellite Working Group (SWG) in summer of 2011. Dawn mission scientists and engineers utilized various computational tools and techniques for Vesta's satellite search. The objectives of this paper are to 1) introduce the natural satellite search problem, 2) present the computational challenges, approaches, and tools used when addressing this problem, and 3) describe applications of various image processing and computational algorithms for performing satellite searches to the electronic imaging and computer science community. Furthermore, we hope that this communication would enable Dawn mission scientists to improve their satellite search algorithms and tools and be better prepared for performing the same investigation in 2015, when the spacecraft is scheduled to approach and orbit the dwarf planet (1) Ceres.

## 1. INTRODUCTION

Planets and dwarf planets follow elliptical paths with the Sun at one of its two foci. *Asteroids*, small bodies in the Solar System primarily located between Mars and Jupiter also orbit the Sun [1]. A *moon* or *natural satellite* orbits a planet, dwarf planet, or asteroid, called its *primary* [2]. Scientists can learn about the origin, early history, and evolution of our solar system by searching for and studying moons of planetary bodies [1-6]. From the orbital characteristics of natural satellites and combining Newton's and Kepler's laws the mass of the primary asteroid can be calculated [7]. This can be done from the Earth based telescopes without the need for in situ spacecraft measurements [4-6]. Scientists also study planetary bodies using robotic spacecraft [7-8]. Protecting the spacecraft from collision with a natural satellite was the primary motivation for the satellite search conducted as the Dawn spacecraft went into orbit around the main belt asteroid (4) Vesta. Secondary motivation was scientific investigation of the asteroid, which we will refer to as Vesta for the remainder of this paper.

The Dawn mission, part of NASA's Discovery program, was launched on September 27, 2007. The mission's primary goals are to study Vesta and dwarf planet (1) Ceres as precursors to planetary bodies. After a four-year, slow and steady approach, the spacecraft finally entered an orbit around Vesta in July 2011, making Dawn the first probe to orbit an object in the Main Asteroid Belt between Mars and Jupiter. Figure 1(a) represents the locations of Vesta and Ceres schematically with the Sun at a focus, the paths of the planets drawn as ellipses and the position of the main belt asteroids at a moment in time between Mars and Jupiter [1]. Figure 1(b) displays the Dawn mission's timeline with

---

respect to the positions of the planets, depicting the path of the spacecraft from its launch in 2007 to entering and exiting Vesta's and Ceres's orbits, and finally the end of the mission in 2015 [8]. Dawn mission's Satellite Working Group utilized various computational tools and techniques for Vesta's satellite search in 2011. The objectives of this paper are to 1) introduce the natural satellite search problem, 2) its computational challenges, and 3) applications of various image processing and computational algorithms for performing satellite search to the electronic imaging and computer science community. We achieve these goals by presenting two case studies, one duplicating a past satellite search performed on Pluto and another on our recent work on Vesta. Furthermore, we hope that this communication would enable Dawn mission scientists to improve their satellite search algorithms and tools to better prepare them for performing the same investigation in 2015, when the spacecraft is scheduled to approach and enter an orbit around the dwarf planet (1) Ceres.

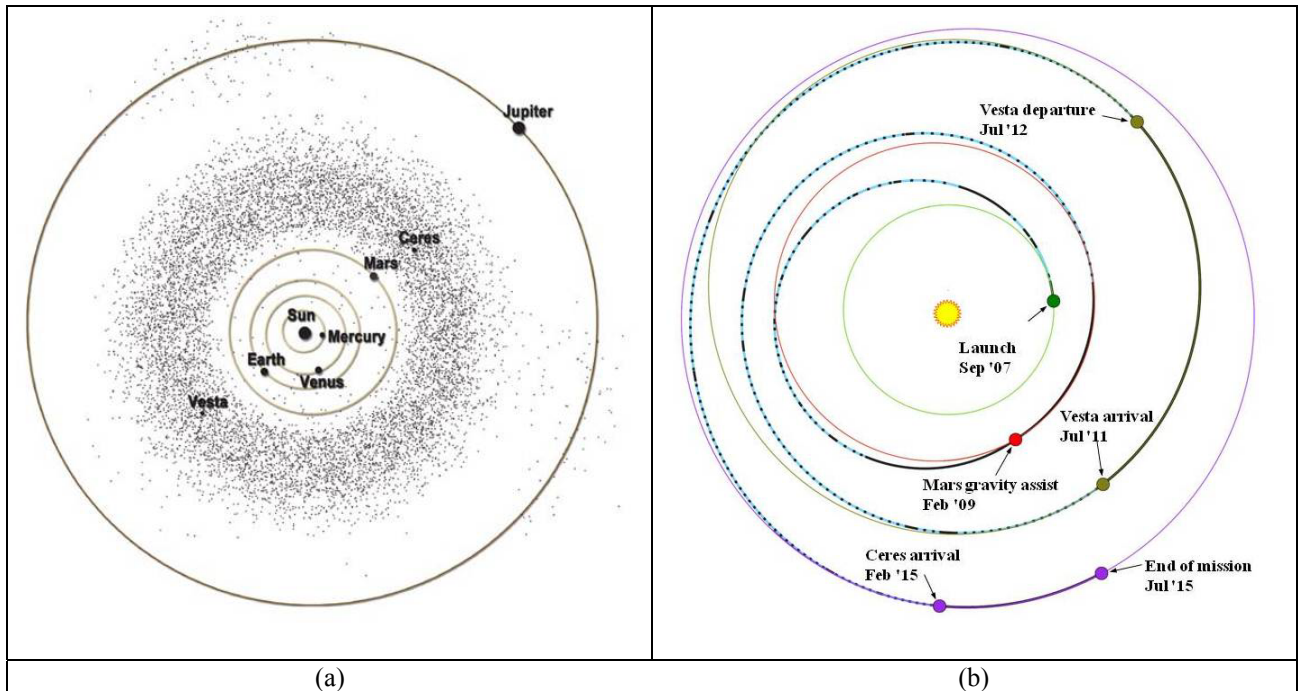

(a) (b)

Figure 1- (a) Plane view of the position of Vesta and Ceres with respect to our solar system and the Main Asteroid Belt between Mars and Jupiter [1]. (b) Dawn mission's timeline from its launch in 2007 to the end of mission in 2015 [8].

The remainder of this paper is organized as follows. First, we introduce the natural satellite search problem in Section 2. Then, in Section 3, we present a brief survey of the past work done for discovery of natural satellites via electronic imaging. We review computational methods, tools, and issues for moon search approaches in Section 4. In particular, we present a case study on verification of the discovery of two of Pluto's moons, Nix and Hydra. Finally, in Section 5 we report on various tools and algorithms that were utilized by Dawn mission's SWG for Vesta's satellite search in 2011. These algorithms include image registration, filtering, and capabilities provided by software packages called astrometry.net [9] and Astrometrica [10]. We will also mention shortcomings of satellite search approaches due to limitations imposed by specific mission designs and science objectives as well as related computational challenges, which could potentially be addressed by the electronic imaging community. Section 6 concludes the paper.

## 2. SATELLITE SEARCH PROBLEM

In this section we describe how astronomers search for moons of planetary bodies or perform a *satellite search*. Moons or satellites of a celestial body are confined to a region of that body's gravitational influence. Scientists approximate the region where the gravitational force of the *primary* body (e.g. Earth, Mars) dominates that of the more massive yet more distant body, the Sun, as a spherical region called the *Hill sphere* [2–6]. When the primary body, with mass $m_p$ rotates around the Sun with mass $M$, the radius of the Hill sphere for the primary body is calculated as

$$r_H = a_p \left(\frac{m_p}{3M}\right)^{\frac{1}{3}} \tag{1}$$

where $a_p$ is the semi-major axis of the primary's orbit around the Sun (e.g. Earth-Sun, Mars-Sun, distance). Considering a mass for Pluto of $1.25 \times 10^{22}$ kilograms and a semi-major axis for its orbit around the Sun and a solar mass of $1.99 \times 10^{30}$ kg, Pluto's Hill sphere is $R_H = 5906.38 \times 10^6 \left(\frac{1.25 \times 10^{22}}{3 \times 1.99 \times 10^{30}}\right)^{\frac{1}{3}} = 7.5561 \times 10^6$ kilometers. Vesta's Hill sphere in May 2007 was measured to be 488 Vesta radii (with Vesta radius being 265 kilometers) or 129,320 kilometers in radius [11], less than two percent of that of Pluto!

Astronomers consider several criteria for validating a potential moon detection as a natural satellite of its primary:
1. It is expected to reside within the primary's Hill sphere.
2. It should be an unknown object and not be listed in any star catalogue.
3. For high resolution data, it should have a point spread function. That is, it should not appear as a spike in only one pixel. Furthermore, it should have the correct point spread function (PSF), one matching that of the observing camera.
4. It should be observed more than once and in consecutive frames.
5. It should obey Newton's and Kepler's laws of motion.

Usually the first three criteria can be verified with a single data set. The fourth criterion requires consecutive observations of the same area. Potential moons that meet the first four criteria are identified. Then, follow-up observations are made at a later time to verify the discovery. That was the case for 2006 Hubble Space Telescope data where follow-up observations of Pluto were made to verify the discovery of two of its moons, Nix and Hydra, that were first observed in 2005.

## 3. LITERATURE SURVEY

Searching for moons around planets has a long history. For example, it was in 1878 when Asaph Hall discovered two moons of Mars which were later named Phobos and Deimos, with the 26-inch Refractor Telescope at the Naval Observatory in Washington D.C. [12-13]. With advances in imaging instruments and technologies as well as robotic exploration of the outer solar system, there has been an increase in satellite search and discoveries around many planets including Jupiter, Saturn, Uranus, Neptune, and Pluto [2-6] with some of the outer planets having more than 50 natural satellites. No additional satellites of Mars have been found.

Pluto and its moons have a fascinating story. It was in 1930 that Clyde Tombaugh discovered Pluto, and then Pluto's largest moon, Charon, was discovered in 1978 [14]. It was much later in 2005 that two smaller moons, Nix and Hydra, were discovered and later verified in 2006 using observations of the Hubble Space Telescope [15-16]. In order for a body to be considered a planet it has to orbit the Sun, have enough gravity to form a spherical shape and enough mass to clear its neighborhood of debris [17]. In other words, it should be the dominant gravitational body in its orbit. Any object, like Pluto, that meets the first two requirements but not the third one is called a dwarf planet. Finally, most recently, in July of 2011, Pluto's fourth moon was discovered using images from the Hubble Space Telescope [18].

Furthermore, asteroids and small bodies beyond Neptune can have moons too [3, 7]. Over two hundred small solar system bodies have companions orbiting them at the time of this writing. The first satellites of asteroids were reported from stellar occultations, where an asteroid passes through the line of sight from the Earth with a star [19]. As the asteroid passes in front of the star, its brightness decreases and stays dimmed until the asteroid passes out of the star's line of sight. Such measurements are used to determine asteroid diameters. In the case of the six asteroid occultations prior to 1980, there was additional stellar dimming observed before and after the main occultation events. Not much credulity was given to these first purported satellite detections until the Galileo spacecraft, on its way to Jupiter, flew past asteroid 243 Ida and imaged a companion in 1993 [20]. Since then all space mission asteroid flybys include a satellite search but no additional moons of any asteroid flyby targets since Ida have been found [21]. Ground based telescopes have detected several asteroid companions such as those around asteroids 45 Eugenia, 762 Pulcova, 90 Antiope, and 87 Sylvia [7]. The search for satellites around asteroid 433 Eros by the Near Earth Asteroid Rendezvous

(NEAR) spacecraft was performed using autonomous search algorithms, and an analysis of the photometric sensitivity of the search indicating a search limited size of 20 meters in diameter. No satellite was detected [7].

With each new technological advance in telescope design and detector capability, planetary astronomers have searched for satellites around the larger and brighter main belt asteroids, including Vesta and Ceres, but to no avail. Gehrels *et al*. [22] conducted the first reported direct imaging search for satellites around large main belt asteroids. Vesta and Ceres were among their targets. An upper search limit of 1-2 km was found for Vesta. Roberts *et al*. [23] used speckle interferometry searching to within 35 – 4000 km from Vesta to a size limit of 51 km. More recently, a satellite search was carried out of Vesta [11] and Ceres [24] using Hubble data. Results of that search indicated no moons of Vesta were found within the region of Vesta's Hill sphere that was searched, with a satellite search limit of 22 +/- 2 meters in radius. No moons of Ceres were detected within a search limit of one kilometer in satellite's size. That the larger asteroids have no moons suggests that the energy of the collisional and dynamical environment exceeded that of the gravitational stability of a natural satellite.

## 4. COMPUTATIONAL ISSUES, METHODS, AND TOOLS

In order to prepare for Dawn's satellite search in July of 2011, we first revisited Hubble Space Telescope (HST) images of Pluto in 2006. With these data we verified the image processing steps necessary for "finding" two of Pluto's moons, Nix and Hydra, that were discovered in 2005 [15-16]. These steps address various issues for satellite search due to motion, image misalignments, and artifacts. In this section, we demonstrate how one can resolve such barriers to satellite search by applying different computational approaches such as image registration and filtering, similar to the work done in [15-16]. We used four processed long exposure HST data files of Pluto. In particular, j9l601m7q_drz.fits, j9l601m9q_drz.fits, j9l601mcq_drz.fits, and j9l601meq_drz.fits files that were retrieved from the Multimission Archive at Space Telescope Science Institute (MAST) [25]. These images have exposure times of 475 seconds and were acquired on February 15, 2006. Figure 2 displays j9l601m7q_drz.fits in two different ways. Figure 2(a) is the image displayed after histogram equalization, while Figure 2(b) is the same image in log scale.

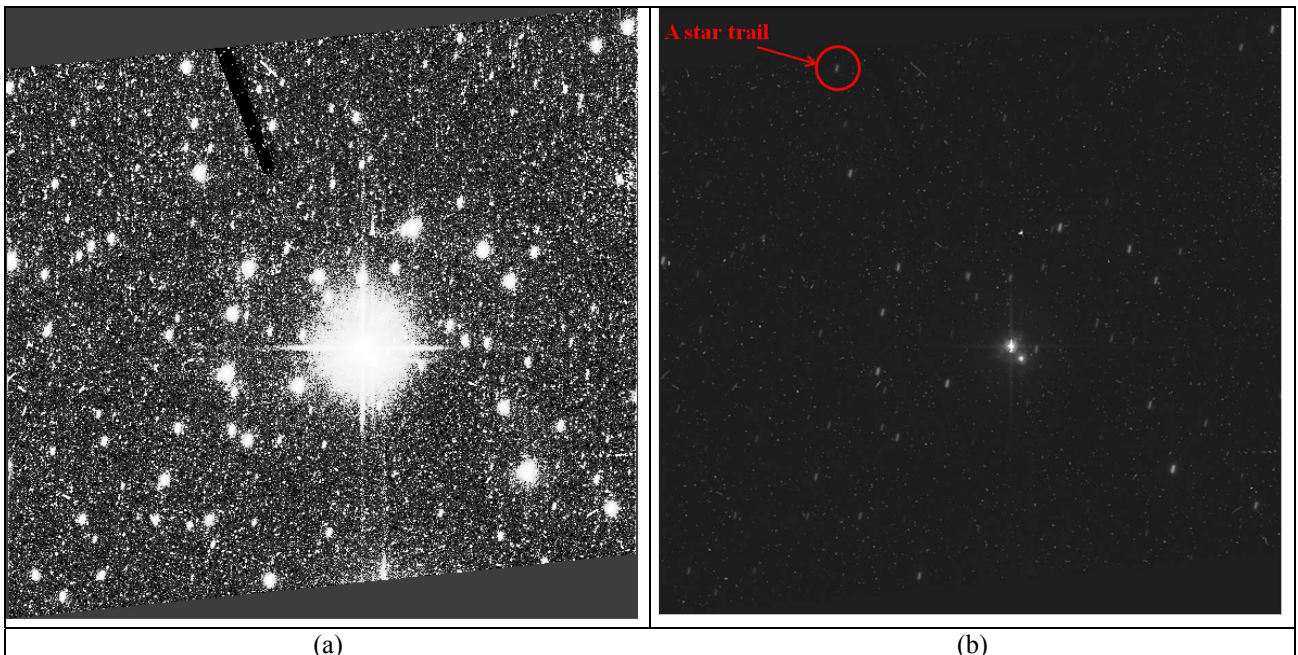

Figure 2: (a) Image j9l601m7q_drz.fits after histogram equalization, (b) Image j9l601m7q_drz.fits in log scale with a marked star trail.

A major issue one should take into account when processing astronomical data is problems due to movement and moving objects. There may be one or more sources of movement when looking at images of the same area of the sky. First, moons, and planetary bodies move over time. Secondly, we often deal with a moving spacecraft resulting in motion parallax and a camera on the spacecraft that has some vibrations referred to as jitter. As a consequence, a pixel with the same coordinates in each frame of the camera does not usually correspond to the same location in the sky. Furthermore, depending on what objects the telescope tracks, we see different effects in the images. If images are obtained when tracking on the fixed background stars, moving objects appear elongated in long exposure images. On the other hand, if a moving object is tracked, then the stars would appear elongated. Figure 3(a) shows sum of the four HST images and the resulting elongation of the background stars when the telescope tracked a moving object, in this case Pluto. We need to perform image registration on different frames of images of the same target (e.g. Pluto, Vesta). Image registration refers to finding the transformation(s) that would align two or more images of the same object or scene that are obtained at different times, from different viewpoints or sensors [26]. We then apply the necessary translation, scaling, or rotation to images such that their corresponding pixels represent the same physical location in the sky and have the same spatial resolution. Pluto and its largest moon Charon reside on the central part of these images, as you can see in Figure 2(b). We used our knowledge of the location of Pluto and Charon to register the image frames with respect to a reference image, image j9l601m7q_drz.fits displayed in Figure 2. After finding the transformation with respect to the reference image, we added the aligned images. Figure 3(b) displays sum of the registered images. As a result, star trails such as those on the lower left corner of Figure 3(a), are better aligned to each other in Figure 3(b).

Next, we have to identify noisy and bad pixels and remove their effects. As you can see in Figure 2(a), the long exposure HST data is very noisy due to the cosmic rays, whose abundance in space is a challenge for astronomical image processing. Cosmic rays are high energy particles coming from sources far away in the galaxy. They usually appear randomly in astronomical images and degrade the quality of the images, with intensity values being orders of magnitude larger than those of the sky. Therefore, astronomers often need multiple images of the same area in the sky captured around the same time for identification and removal of the cosmic rays. We stacked the four registered image frames (that were registered using Pluto and Charon as features), and evaluated each pixel along the direction of the stack. We applied a median filter to each four-dimensional pixel value, and in doing so eliminated most of the cosmic rays, and finally we added all four registered and median-filtered frames. Figure 3(c) displays the result of this process. On the lower right quadrant of the image, towards the middle, we can see Nix and Hydra within a marked box. Please note that Figures 3(a-c) are displayed in inverted grayscale after applying histogram equalization. Much cleaner and better processed results were published in [15-16] as displayed in Figure 4 (the celestial North direction is toward the right). A more statistically robust approach for removing cosmic rays, and also in this case star trails, is to apply the *resistant mean* algorithm, also referred to as *trimmed mean* algorithm. Details of this algorithm, often used by NASA scientists, are available at [27] in the commercial IDL language [28].

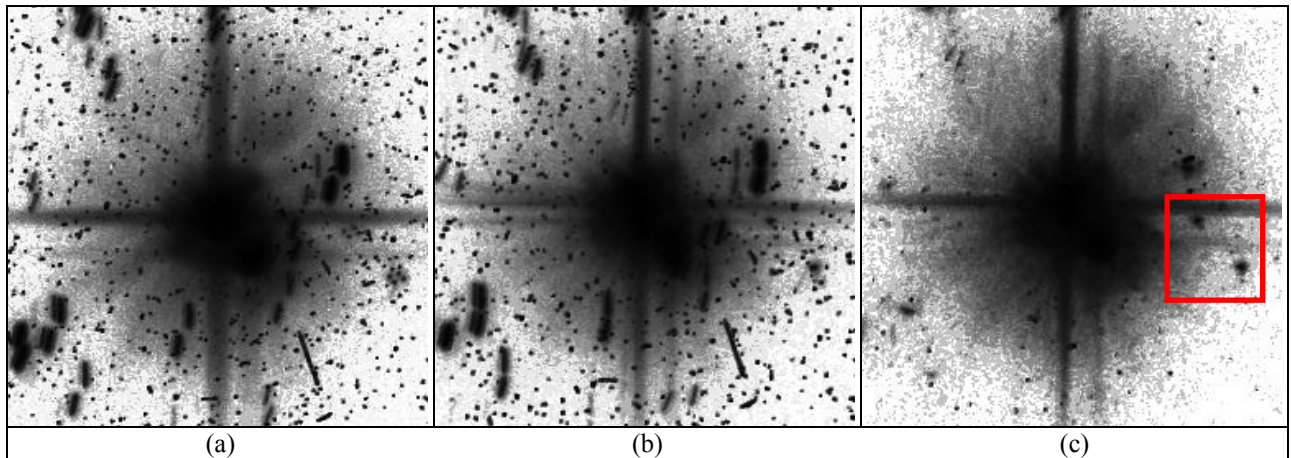

(a) (b) (c)

Figure 3: processed images of the long exposure HST observations of Pluto captured on February 15, 2006, in inverted grayscale values after applying histogram equalization. (a) Sum of the original frames, (b) sum of the registered frames, and (c) sum of the registered frames after applying the median filter, Nix and Hydra reside in the marked box.

Other issues to consider include false detections such as spacecraft debris, systematic artifacts such as dead or hot pixels, and calculating the limits to brightness and detectable motion for the moon search. These limits are defined by several factors including the camera resolution, distance of the camera from the observed targets, and the limiting detection

*magnitude*s [11], or astronomical brightness units, of the observations when drawing conclusions about the search results. SWG continues its work determining Dawn's satellite search limits for Vesta.

Finally, it is important not to confuse the identified moon candidates with other known or unknown celestial objects such as asteroids. For this purpose astronomers benefit from the existing sky catalogs and find the coordinates of the moon candidates in those catalogs. However, this might not always be possible due to the images' resolution or field-of-view limitations. This is where a powerful computational tool, astrometry.net, plays an important role [9]. We used this tool extensively when performing satellite search for the Dawn mission. We elaborate on this tool as well as another one called Astrometrica [10] in the next section.

Ground-based telescopes are also often used for detecting moving objects using similar methods [4-6]. The same region of the sky is observed at least three times. Registered images are processed for removing systematic artifacts, identifying the fixed stars and moving objects. Then, astronomers search in sky catalogs for positions of the moving object candidates. Candidates that were not listed in catalogs and also move in the same direction with a fixed speed from frame to frame are then identified. Future observations are then planned to verify the motion and to calculate the trajectory of the identified moving objects.

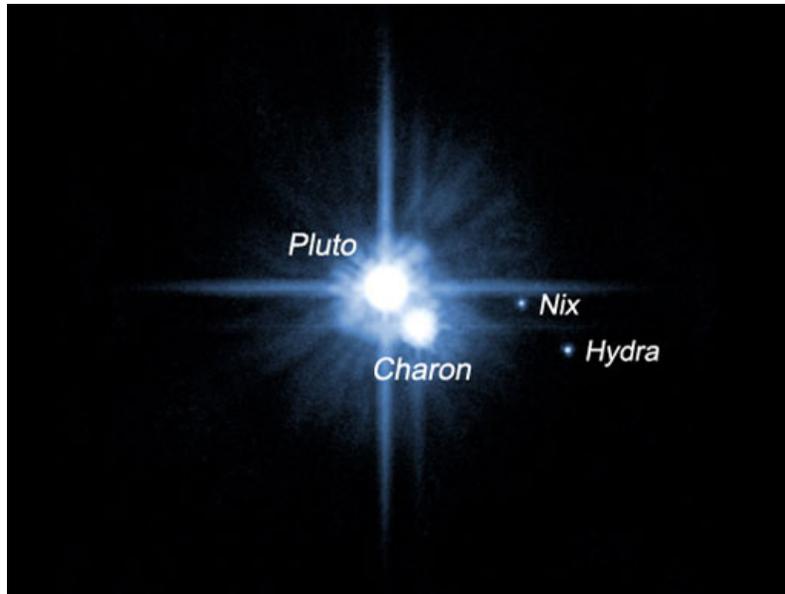

Figure 4: Pluto and its moons from observations made by the Hubble Space Telescope. Image Credit: Stern *et. al* [15-16] North direction is towards the right, and East is towards the top of the image.

## 5. DAWN MISSION'S SATELLITE SEARCH FOR ASTEROID VESTA

In this section we report on the Dawn mission's Satellite Working Group (SWG) efforts in 2011. SWG was tasked with finding moons of Vesta primarily for spacecraft safety. Scientific discovery of moons of Vesta, while not Dawn mission's objective, naturally was a byproduct of this investigation. It was an opportunity that Dawn scientists did not want to miss for better understanding the dynamical history of the asteroid Vesta. Figure 5(a) displays Hubble observations of Vesta in 2007 and Ceres in 2004 from a distance of more than 176.5 and 246.8 million kilometers respectively. The Dawn mission has enabled us to observe and study Vesta from a much closer distance, and will do the same for Ceres in 2015. Figure 5(b) displays a Dawn image of Vesta from a distance of 5,200 kilometers on July 22, 2011. In this section, we first describe datasets that were used for satellite search of Vesta by Dawn's SWG in 2011. We then present the computational algorithms and tools that we benefited from. Finally, we report our results.

### 5.1 Data sets

Dawn images that were used for the satellite search of Vesta were captured via its Framing Camera-2 (FC-2), the German contribution to the Dawn mission [29]. These images were categorized into two groups: Optical Navigation

(OpNav) and Satellite Search images. Each of these two-dimensional images had 1024 × 1024 pixels, of 2-byte integer values, resulting in 2 megabytes of data. OpNav images for this work were obtained during May 3 to July 18 of 2011, as the spacecraft was approaching asteroid Vesta and finally entered its orbit in July, 2011. The distance of the Dawn spacecraft from asteroid Vesta in these OpNav images ranged from 10,500 to 1,217,900 kilometers. OpNav images had exposure times ranging from 8 to 1500 milliseconds. These images were obtained while the Dawn spacecraft was pointing at Vesta. Therefore, in the summed image of the consecutive aligned frames, as well as in long exposure images, we could expect all stars to be trailed in one direction. Streaks that appeared in different direction from the stars could either be cosmic rays, asteroids, or possibly a moving satellite.

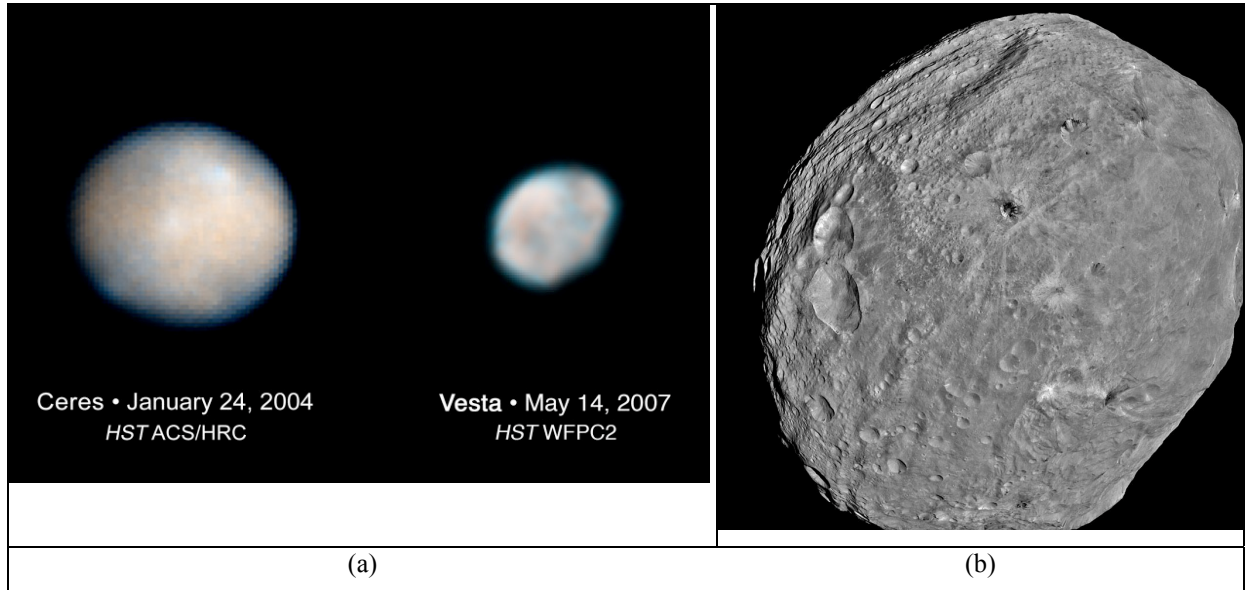

Figure 5-(a) Hubble Space Telescope images of dwarf planet Ceres via ACS/HRC camera on January 24, 2004 and of asteroid (4) Vesta on May 14, 2007 via WFPC2 camera. Credit: NASA/ESA, J. Parker (Southwest Research Institute), and Lucy McFadden (University of Maryland), STScI-PRC07-27a. (b) NASA's Dawn spacecraft obtained this image of the asteroid Vesta with its framing camera on July 24, 2011. Credit: NASA, JPL-Caltech, UCLA, MPS, DLR, IDA", http://dawn.jpl.nasa.gov.

Satellite Search images were obtained on July 9-10, 2011, when the Dawn spacecraft executed a dedicated satellite search sequence. This sequence consisted of three mosaics of six stations surrounding Vesta covering most of Dawn's region of operation. Figure 6(a) displays the mosaic design for data acquisition via six pointing stations. The inner circle in this figure has a radius of 475 kilometers around Vesta, while the middle and outer circles have radii of 950 and 3000 kilometers respectively representing the planned orbiting altitudes of the Dawn spacecraft. Each station consists of four sets of three images with exposure times of 1.5, 20 and 270 seconds separated by five second intervals, as Figure 6(b) demonstrates. In other words, for each of the six stations, we obtained images with three different exposure times, repeated four times. This process was performed three times at different time intervals when collecting Satellite Search data. Different integration times were used to accommodate different amounts of scattered light from Vesta and to search to different levels of detection and therefore size. We searched each station for moving objects within the time frame of the mosaics.

### 5.2 Computational approaches and tools

The Dawn mission's SWG applied different approaches for satellite search to the data sets mentioned in Section 5.1. These approaches included visual detection of moving objects by astronomers, automatic object detection, and image subtraction. The team also benefited from open-source computational software packages such as astrometry.net [9] and Astrometrica [10]. These tools greatly helped with the quick turnaround of the results, which was of high importance and necessity for the Dawn spacecraft's safety. For all approaches, we started working with raw OpNav and Satellite Search data of the FC-2 [29].

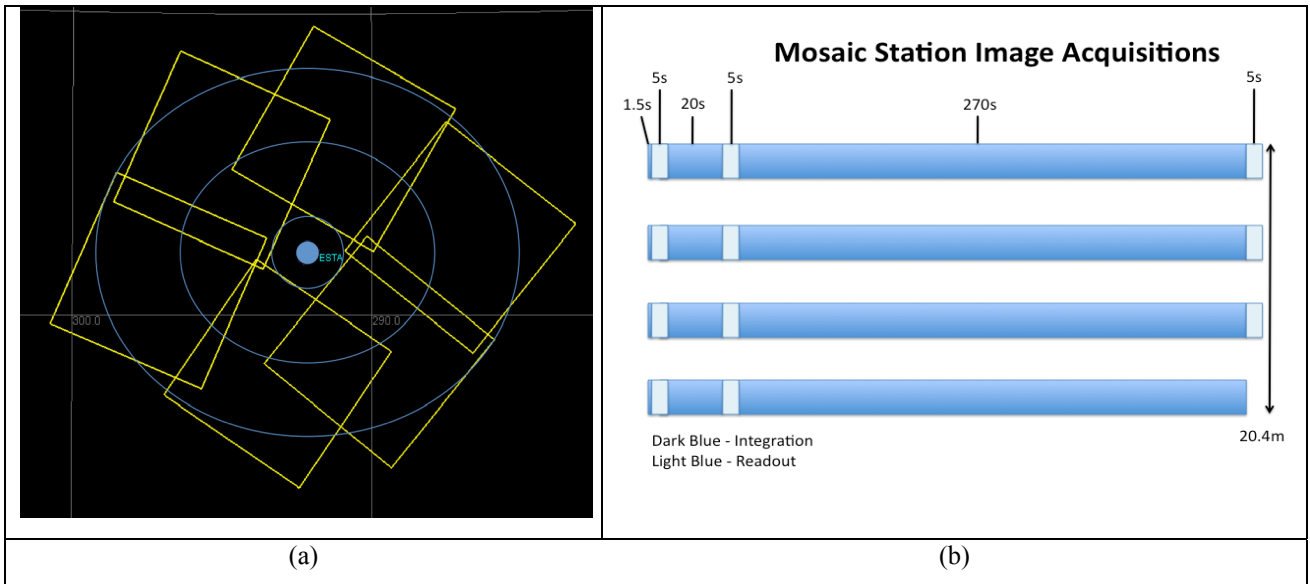

|                (a)                |                (b)                |

Figure 6-(a) station design for Dawn's satellite search of Vesta on July 9-10, 2011, (b) image acquisition exposure and interval times.

### 5.2.1 Visual inspection

This approach heavily relied on identifying moving objects that moved in a different direction than the background stars in long exposure data or when comparing consecutive image frames of the same area in the sky with the same exposure times. These images were stacked in chronological order that they were captured and then registered to each other. Images were aligned using Astrometrica [10], a software package that reads images in the standard astronomy format, FITS [30], and aligns them by searching for pairs of stars of the same distances, position angles, and differences in brightness. After building a set of pairs of stars in the reference image, it searches for possible matches in other images and calculates the necessary transformations for their alignments. Then, frames in such stacks were compared and blinked from one to the next, aiming to detect moving satellites by expert eyes. In order to have a more informed search, we calculated the range of expected motions for possible satellites. Also, positions of the known asteroids were transformed by Astrometrica's author, Herbert Raab [10], to Vesta-centric frames that were being processed. Those OpNav images that were taken from a distance of 225,600 kilometers or less, and all Satellite Search data except those with the very short exposure time of 1.5 seconds were processed this way. No moons were discovered with this approach, but asteroids 972 Cohnia, displayed in Figure 7(a), was "found" and observed this way.

Similarly, after stacking, registering, and adding frames with the same exposure times, stars in OpNav images appeared as trails of the same direction and length. Figure 7(b) displays one of these products, showing the motion of Vesta relative to the background stars, after applying an 11 boxcar filter for removing the scattered light from Vesta. Red circles represent orbits of different distance from Vesta to guide the eye. We searched for trails of different direction and length from those of the background stars in such OpNav products. We then further examined those trails to remove other moving objects such as background asteroids. A few previously known asteroids were identified this way, but no moons were found.

### 5.2.2 Automated object-detection

One of the most systematic and automated approaches we used for satellite search was based on automatic object detection. The first crucial step for this approach was to determine where on the celestial sphere the camera was looking at in order to rule out the possibility of mistaking known celestial bodies, such as stars or asteroids, as newly discovered objects. We used astronomy.net for this task [9]. Astrometry.net is open source software that was developed with partial support from the National Science Foundation (NSF), National Aeronautics and Space Administration (NASA), and the Canadian National Science and Engineering Research Council. This software receives sky images as input and returns the World Coordinate System (WCS) of the images [9]. WCS is a standard description of the nonlinear transformation between image coordinates and the sky coordinates, and is provided in the form of standard metadata which can be included in FITS data files, a standard for astronomy images [30]. Astrometry.net software relies on patterns of the known stars, asteroids, etc. for identifying celestial bodies and calculating their coordinates. It achieves this goal via an efficient pattern matching algorithm for points in the two-dimensional space. Such patterns can have unknown locations, scales, and orientation. The software compares patterns in an image against a collection of known sky patterns in its database. This database is a collection of so called *index* files, which should be installed as part of the software package. When the software successfully matches a pattern in an input image with one in its database, the image is *solved*. Astrometry.net indices were built based on United States Naval Observatory-B (USNO-B) [31] catalog. For a detailed description of how index files were created and the methods that are used for pattern matching please see [9]. One limitation of this software however, is that it can only take into account the effect of the geometric projection of the instrument detector plane on to the celestial sphere, and most instruments have distortions which are not accounted for, reducing the precision of the solution. In order for the software to work properly, it is essential to have enough stars in each image's field-of-view. This was the case until the Dawn spacecraft got closer than 42000 kilometers to asteroid Vesta and later entered its orbit, when approximately 30 percent or more of the field-of-view was filled by the asteroid itself. For images that did not have enough stars in their field-of-view, the software took either too long to find a solution (a pattern match) or could not solve the image completely.

The next step was to automatically locate objects in the individual image frames. This was done using the FIND procedure in the IDL [28] astronomy library, which is based on the classic DAOPHOT algorithm [32]. The sky backgrounds were determined using the related MMM (Mean,Median,Mode) procedure in this library. Essentially, this algorithm locates positive brightness perturbations above the sky background in an image and computes parameters related to how well the object is described by a Gaussian point spread function. These parameters can be used to identify point sources (stars), extended sources (galaxies, resolved asteroids or moons), and image defects/artifacts.

After identifying enough point sources in each frame, we used the approximate WCS coordinates of each object to identify them in both the UCAC3 [33] and GSC2 [34] astrometric reference catalogs which are based on the International Celestial Reference System (ICRS) [35]. This allowed us to recalibrate the astrometry using a higher-order polynomial solution to model the instrument focal plane which reduced the errors in the coordinates from around 1 arcminute to 1.5-2.0 arcseconds.

By comparing the object detections for all frames in each observation series, we can remove detections which are cosmic ray hits since they only appear in single frames. Similarly, we can remove detections due to hot or defective pixels since they don't move with the star background. Since the detector's focal plane is under sampled, the apparent brightness of the individual objects can vary as they move across the detector. Therefore, real stars in the field near the faint detection limit are sometimes not visible. Consequently, we adopted a criterion that 'real' objects would be visible in at least 50% of the frames in any observation series.

At the end of the processing we have a list of observations of linked objects. The very small number of objects which did not match a known background reference star were individually examined to see if they were potential moons or asteroids but all were rejected as such candidates.

Figure 7(c) displays the results of this approach on an OpNav image with detected objects circled. Green circles are objects that could be identified and matched with objects in a sky catalogs and red circles are objects that could not be matched with known objects in selected sky catalogs. No moons were identified.

### 5.2.3 Image subtraction

Finally, we used a simple and quick approach for identifying potential satellites that involved image subtraction. This procedure could be used only for some OpNav images when the spacecraft did not change its planar orientation. Thus, this final approach was not used for Satellite Search images. The first three images in each OpNav series were selected and stacked. These images were well-aligned to each other and we did not apply any registration to them.

Then, for each pixel in the stack, the maximum of the three values was ignored to quickly remove cosmic rays. An image frame was created by averaging the remaining two values for each pixel. Similarly, another frame was created with the last three OpNav images in the same series. Subtracting one of these frames from the other, results in the bright and dark spots that point to a very specific direction, which is the direction of the spacecraft's motion. A variation from this direction will imply a moving object candidate. Figure 7(d) shows one product of this approach. A moon would have a different direction and show a different separation compared to the background stars, assuming its brightness did not change significantly between the observation intervals. No moving object was observed this way other than occasionally an asteroid.

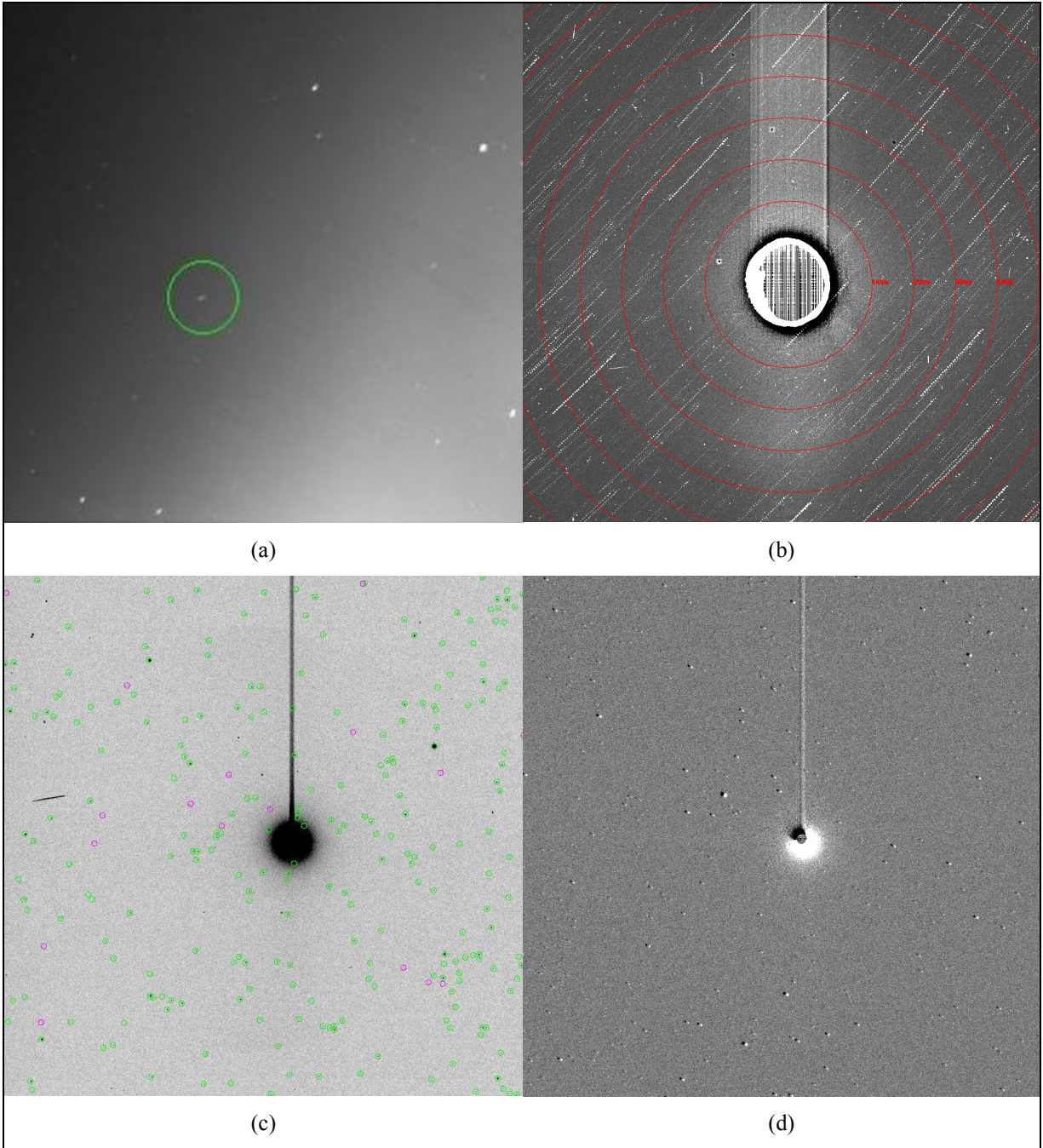

Figure 7- (a) A background asteroid with different direction of motion relative to background stars. This is asteroid 972 Cohnia found in the mosaic search sequence. (b) Registered and coadded frames to search for trails of different direction and length from the background stars. Vesta is in the center of the frame, the vertical streak is smeared residual signal from Vesta due to the detector's readout characteristics. (c) The result of the automatic object detection approach for an OpNav image, and (d) the result of the "Image subtraction" approach on an OpNav image.

## 5.3 Results

We applied several techniques to OpNav and Satellite Search images acquired by the Dawn spacecraft to search for moons of the asteroid Vesta in summer of 2011. We searched for objects that moved in different directions from the stars in image frames. We also searched for detected objects that were not identified in existing star catalogues. We applied image registration and object detection algorithms to our images. The visual inspection approach had an advantage for later OpNav images obtained from a closer distance to Vesta over the automatic object detection approach. As asteroid Vesta occupied more than 30 percent of the field of view, there were not enough background stars to help solve the images with astrometry.net, to obtain their WCS coordinates, and to benefit from automatic object detection algorithm. Similarly, the automatic object detection approach had an advantage over the visual inspection approach for earlier OpNav images, when objects were too far or too faint to be easily detected by eye. The image subtraction method could only be applied to a subset of OpNav images, when the spacecraft did not change its orientation. We also benefited from several open-source computational software packages. WCS coordinates of the images were determined using astrometry.net software [9]. Images were aligned and blinked using Astrometrica [10]. Celestial objects were detected using DAOPHOT [32] algorithm. Background stars were compared with star catalogs and unidentified objects were noted. We have found no evidence of objects in motion near Vesta to date. Filtered images also have shown no evidence of any object in motion other than a few known asteroids that happen to be in the field of view.

## 6. CONCLUSION

In this paper we introduced the natural satellite, or moon, search problem to the electronic imaging community. We presented computational challenges, approaches, and existing tools for solving it. Furthermore, we presented two case studies for satellite searches in detail. First, we reviewed computational approaches for verifying the discovery of Pluto's moons, Nix and Hydra, using Hubble data of Pluto in 2006. Second, we described the work of the Dawn mission's Satellite Working Group (SWG) in summer of 2011, searching for moons of the asteroid Vesta. We employed three independent approaches successfully and concluded that there are no natural satellites, moons, of Vesta in the portion of the Hill sphere of Vesta that we searched. This paper was a preliminary report on methods used for finding moons of Vesta. A future report by the SWG will provide the analysis on various limits on this search such as magnitude and size, as well as motion. A satellite search will also be made for dwarf planet Ceres as the Dawn spacecraft approaches it in 2015 for spacecraft safety as well as possible scientific discovery. Therefore, applying the lessons learned and having more advanced and efficient computational tools and processes for satellite search continues to be of high importance and value to planetary scientists, astronomers, and missions' optical navigation specialists. We hope that this communication with the computational community better prepares us for such future investigations.

## ACKNOWLEDGEMENTS

We acknowledge the contributions of our colleagues in the Satellite Working Group: Mark V. Sykes, Pasquale Tricarico, and David O'Brien (Planetary Science Institute, Tucson, Arizona); Pablo Gutierrez-Marques, Stefan Schroeder, Andreas Nathues, Horst U. Keller, and Holger Sierks (Max-Planck-Institut für Sonnensystemforschung, Lindau, Germany); Christopher T. Russell and Steve Joy (IGPP, UCLA); Jian-Yang Li (University of Maryland); Stefano Mottola (DLR, Berlin); Robert A. Jacobson, Carol A. Polanskey, Marc D. Rayman, Carol A. Raymond, and Stacy Weinstein-Weiss (Jet Propulsion Laboratory, California Institute of Technology).

The Dawn mission to Vesta and Ceres is managed by the Jet Propulsion Laboratory, for NASA's Science Mission Directorate, Washington, D.C. It is a project of the Discovery Program managed by NASA's Marshall Space Flight Center, Huntsville, Alabama. UCLA is responsible for overall Dawn mission science. Orbital Sciences Corporation of Dulles, Virginia, designed and built the Dawn spacecraft. The German Aerospace Center, the Max Planck Institute for Solar System Research, the Italian Space Agency, and the Italian National Astrophysical Institute are part of the mission team.

## REFERENCES

[1] Britt, D. T., Consolmagno, G. and Lebofsky, L., "Main-Belt Asteroids", Encyclopedia of the Solar System, 2nd edition. Eds. McFadden, L. A., Weissman, P. R. and Johnson, T. V., Elsevier/Academic Press, New York, 349-364 (2007).
[2] Buratti B. J. and Thomas, P. C., "Planetary Satellites" in Encyclopedia of the Solar System, 2nd edition. Eds. McFadden, L. A., Weissman, P. R. and Johnson, T. V., Elsevier/Academic Press, New York, 365-382 (2007).
[3] Merline, W. J., Weidenschilling, S. J., Durda, D. D., Margot, J-L., Pravec, P. and Storrs, A. D., "Asteroids Do Have Satellites", in Asteroids III, The University of Arizona Press, 289-313 (2002).
[4] Sheppard, S. S., Jewitt, D. and Kleyna, J., "A survey for outer satellites of Mars: Limits to completeness", The Astronomical Journal 128(5), 2542–2546 ( 2004).

[5] Sheppard, S. S., Jewitt, D. and Kleyna, J., "An ultradeep survey for irregular satellites of Uranus: Limits to completeness," The Astronomical Journal 129(1), 518–525 ( 2005).
[6] Sheppard, S. S., Jewitt, D. and Kleyna, J., "A survey for "normal" irregular satellites around Neptune: Limits to completeness," The Astronomical Journal 132(1), 171–176 ( 2006).
[7] Merline W. J., Tamblyn, P., Chapman, C. R., Colwell, W. B., Gor V., Burl M. C., Bierhaus E. B. and Robinson, M. S., "An Autonomous Search for Moons during Approach of the NEAR Spacecraft to Asteroid Eros", Proc. 6th International Symposium on Artificial Intelligence and Robotics & Automation in Space ( i-SAIRAS), Canadian Space Agency, St.-Hubert, Quebec, Canada, June 18-22, (2001).
[8] Rayman,  M. D. and Mase, R. A., "The second year of Dawn mission operations: Mars gravity assist and onward to Vesta", *Acta Astronautica*, 67**,** 483-488 (2010).
[9] Lang, D., Hogg, D. W., Mierle, K., Blanton, M. and Roweis, S., "Astrometry.net: Blind astrometric calibration of arbitrary astronomical images", The Astronomical Journal 137, 1782–1800 (2010).
[10] Herb, R., "Astrometrica", http://www.astrometrica.at , Accessed November 7, 2011.
[11] McFadden, L. A., Bastien, F. A., Crow, C. A., Hamilton, D. P., Li, J. and Mutchler, M. J., "Search for Satellites of Vesta: Upper Limits on Size", American Astronomical Society, DPS meeting  41, 53.06 (2009).
[12] Hall, A., [Observations and orbits of the satellites of Mars: With data for ephemerides in 1879]*,* Government Printing Office, Washington D. C. (1878).
[13] Hall, A., "Observations and orbits of the satellites of mars: With data for ephemerides in 1879". Astronomical and Meteorological Observations made at the U.S. Naval Observatory, 15, d1–d46 (1878).
[14] Christy, J. W. and Harrington, R. S., "The discovery and orbit of Charon", Icarus 44(1), 38-40 (1980).
[15] Stern, S. A., Mutchler, M. J., Weaver, H. A. and Steffl, A. J., "The Positions, Colors, and Photometric Variability of Pluto's Small Satellites from HST Observations: 2005-2006",  proceedings of the 38th Lunar and Planetary Science Conference, Lunar and Planetary Institute (LPI) Contribution, 1338, 1722 (2007).
[16] Weaver, H. A., Stern, S. A., Mutchler, M. J., Steffl, A. J., Buie, M. W., Merline, W. J., Spencer, J. R., Young, E. F. and Young, L. A., "Discovery of two new satellites of Pluto". *Nature*,  439(7079), 943–945 (2006).
[17] Cain, F., "Why Pluto is no longer a planet", 2008. http://www.universetoday.com/13573/
[18] Showalter, M. R., Hamilton, D. P., Stern, S. A., Weaver, H. A., Steffl, A. J. and Young, L. A., "New Satellite of (134340) Pluto: S/2011 (134340)", IAU Circ., 9221(1), (2011).  Edited by Green, D. W. E.
[19] Maley, P. D., "In Search of Satellites of Minor Planets", Royal Astronomical Society of Canada, 74(Dec.), 327-333 (1980).
[20] Chapman, C. R., Veverka, J., Thomas, P. C., Klaasen, K., Belton, M. J. S., Harch,  A., McEwen, A., Johnson, T. V., Helfenstein, P., Davies, M. E.,  Merline, W. J. and Denk, T., "Discovery and Physical Properties of Dactyl, a satellite of asteroid 243 Ida.", Nature, 374,  783-785 (1995).
[21] Johnston, R. W., "Asteroids with Satellites", http://www.johnstonsarchive.net/astro/asteroidmoons.html, Accessed Dec. 2011.
[22] Gehrels, T., Drummond, J. D. and Levenson, N. A., "The Absence of Satellites of Asteroids", Icarus, 70, 257-263.
[23] Roberts, L. C., McAlister, H. A. and Hartkoff, W. I., "Speckle Interferometric Survey for Asteroid Duplicity," The Astronomical Journal,  110(5), 2463-2468 (1995).
[24] Bieryla, A., Parker, J. W., Young, E. F., McFadden, L. A., Russell, C. T., Stern, S. A., Sykes, M. V. and Gladman, B., "A Search for Satellites around Ceres", The Astronomical Journal, 141(6), 197 (2011).
[25] The Multimission Archive at STScI. http://archive.stsci.edu/.  Accessed: November 2, 2011.
[26] Le Moigne, J., Netanyahu, N. S. and Eastman, R. D., Eds., [Image Registration for Remote Sensing], Cambridge University Press, Cambridge, United Kingdom, (2011).
[27] Freudenreich, H., Resistant mean program in IDL. http://idlastro.gsfc.nasa.gov/ftp/pro/robust/resistant_mean.pro, (1989). Accessed: June 3, 2011.
[28] Interactive Data Language (IDL). http://www.ittvis.com/language/en-US/ProductsServices/IDL.aspx. Accessed: June 3, 2011.
[29] Sierks, H., Keller, H. U., Jaumann, R., Michalik, H., Behnke, T., Bubenhagen, F., Büttner, I., Carsenty, U., Christensen, U., Enge, R., Fiethe, B., Gutiérrez Marqués, P., Hartwig, H., Krüger, H., Kühne, W., Maue, T., Mottola, S., Nathues, A., Reiche, K.-U., Richards, M., Roatsch, T., Schröder, S., Szemerey, I. and Tschentscher, M., "The Dawn Framing Camera",  Space Science Reviews, 1-65(2011)
[30] Wells, D. C., Greisen, E. W. and Harten, R. H., "FITS - a flexible image transport system", Astronomy and Astrophysics Supplement Series, 44, 363–370 (1981).
[31] Monet, D. G., *et al*. "The USNO-B Catalog", The Astronomical Journal, 125(2), 984-993(2003).
[32] Stetson, P. B., "DAOPHOT - A computer program for crowded-field stellar photometry", Publications of the Astronomical Society of the Pacific, 99, 191-222(1987).
[33] Zacharias, N., *et al*. "The Third US Naval Observatory CCD Astrograph Catalog (UCAC3)", The Astronomical Journal, 139(6), 2184-2199 (2010).
[34] Lasker, B., *et al.,* "The Second-Generation Guide Star Catalog: Description and Properties", The Astronomical Journal, 136(2), 735-766 (2008).
[35] Feissel, M. and Mignard, F., "The Adoption of ICRS on 1 January 1998: Meaning and Consequences", Astronomy and Astrophysics, 331, L33-L36 (1998).